\documentstyle[11pt,epsfig]{article}

\begin{document}

\title{Decay Constants of Heavy Meson of $0^-$ State \\
in Relativistic Salpeter Method}
\vspace{10mm}

\author{G. Cveti\v c$^1$\footnote{gorazd.cvetic@usm.cl},~~
C.~ S.~ Kim$^2$\footnote{cskim@yonsei.ac.kr},~~
Guo-Li~ Wang$^2$\footnote{glwang@cskim.yonsei.ac.kr}~~ and ~~
Wuk Namgung$^3$\footnote{ngw@dgu.ac.kr} \\
\\
{\it \small $^1$ Dept.~of Physics, Univ.~T\'ecnica
Federico Santa Mar\'{\i}a, Valpara\'{\i}so, Chile}\\
{\it \small $^2$ Department of Physics, Yonsei University, Seoul
120-749, Korea}\\
{\it \small $^3$ Department of Physics, Dongguk University, Seoul 100-715, Korea} }
\date{}
\maketitle

\baselineskip=24pt
%\thicklines
\begin{quotation}
\vspace*{1.5cm}
\begin{center}
  \begin{bf}
  ABSTRACT
  \end{bf}
\end{center}

\vspace*{0.5cm}
\noindent The decay constants of pseudoscalar
heavy mesons of $0^-$ state are computed by means of the
relativistic (instantaneous) Salpeter equation.
We solved the full Salpeter equation without making any further approximation,
such as ignoring the small component wave function.
Therefore, our results for the decay constants
include the complete relativistic contributions from the light and the heavy
quarks. We obtain $F_{D_s} \approx~ 248~\pm ~27~$,
$F_{D} \approx~ 230~\pm ~25~ (D^0,D^\pm)$,
$F_{B_s} \approx~ 216~\pm ~32~$,
$F_{B} \approx~ 196~\pm ~29~ (B^0,B^\pm)$,
$F_{B_c} \approx~ 322~\pm ~42~$
and $F_{\eta_c} \approx~ 292~ \pm ~25~$ MeV.
\end{quotation}

\newpage
 \setcounter{page}{1}
\section{Introduction}

The decay constants of mesons are very important quantities.
The study of the decay constants has become an interesting
topic in recent years, since they provide a direct source of
information on the Cabibbo-Kobayashi-Maskawa matrix elements.
In the leptonic or nonleptonic weak decays of $B$ or $D$
mesons, the decay constants play an important role.
Further, the decay constant plays an essential role in the
neutral $D-\bar{D}$ or $B-\bar{B}$ mixing process.

Up until now, the only experimentally obtained values of the decay
constants are those of $F_{D^{+}}$ and $F_{D_s}$. The first value
is $F_{D^{+}}=300^{+180+80}_{-150-40}$ MeV by BES~\cite{fp}, with
very large uncertainties. The experimental values of $F_{D_s}$
have been obtained from both $D_s \to \mu\nu_{\mu}$ and $D_s \to
\tau\nu_{\tau}$ branching fractions by many experimental
collaborations
(Refs.~\cite{wa75,cleo1,e653,l3,delphi,bes,cleo2,beatrice,opal,aleph}).
They are shown in Table 1. The central values from various
experiments range from 194 to 430 MeV. The experimental
uncertainties in each experiment are large, even in the most
recent measurement, by ALEPH~\cite{aleph} ($F_{D_s}=285\pm19\pm40$
MeV), which has the smallest uncertainty. Further, also in ALEPH's
measurement, the contribution from the decay $D_s \to
\mu\nu_{\mu}\gamma$ is ignored. Unlike the tree level case which
is helicity-suppressed, this radiative decay does not
have the helicity suppression. Therefore, this radiative decay may
contribute several per cent to the branching ratio \cite{wang},
and may thus cause a sizeable change in the value of the decay
constant $F_{D_s}$. Fortunately, new experiments such as Belle,
BaBar, Tevatron Run II and CLEO-c will give us a wealth of
precision data for $B$ and $D$ mesons soon, and will determine the
decay constants to a few per cent.

Many theoretical groups are working on the calculation
of the decay constants, using different models, for
example, lattice QCD, QCD sum rules, and the potential model.
In Fig.~1 (taken from Ref.~\cite{becirevic}), as an example,
the world average of the quenched lattice results for
$F_{B_s}$ \cite{world,ryan} is shown. From
this Figure, we conclude that they give a stable
estimate of the decay constant $F_{B_s}$ over several years,
but the uncertainties are not small and still remain unchanged over the
last several years of work. More precise predictions are
still not available.
Decay constants of other mesons
calculated by lattice methods face the same problem as
$F_{B_s}$, the uncertainties are still large. Precise
experimental results with uncertainties of only a few per cent
will be obtained soon. Therefore, advancing the lattice QCD
calculations is an urgent task in the future.
Calculations with different models are and will continue to be
needed, as a means to cross-check the lattice results.

\begin{table}[hbt]\begin{center}
\caption{Summary of the experimental determinations of the decay
constant $F_{D_s}$.} \vspace{0.5cm}
\begin{tabular}
{|c|c|}\hline Ref.&$F_{D_s}$ (MeV)\\\hline \cite{wa75} WA75
'93&232$\pm$45$\pm$52\\\hline \cite{cleo1} CLEO I
'94&344$\pm$37$\pm$52$\pm$42
\\\hline \cite{e653} E653 '96&194$\pm$35$\pm$20$\pm$14
\\\hline \cite{l3} L3 '97&309$\pm$58$\pm$33$\pm$38
\\\hline \cite{delphi} DELPHI '97&330$\pm$95\\\hline \cite{bes}
 BES '98&430$^{+150}_{-130}$$\pm$40
\\\hline \cite{cleo2} CLEO II '98
&280$\pm$19$\pm$28$\pm$34
\\\hline \cite{beatrice} BEATRICE '00&323$\pm$44$\pm$12$\pm$34
\\\hline \cite{opal} OPAL '01&286$\pm$44$\pm$41
\\\hline \cite{aleph} ALEPH '02&285$\pm$19$\pm$40
\\\hline
\end{tabular}
\end{center}
\end{table}

\begin{figure}[tbh]
\vspace*{-3mm}\centerline{\hspace*{-5.5mm}\epsfxsize=0.5\textwidth\epsffile{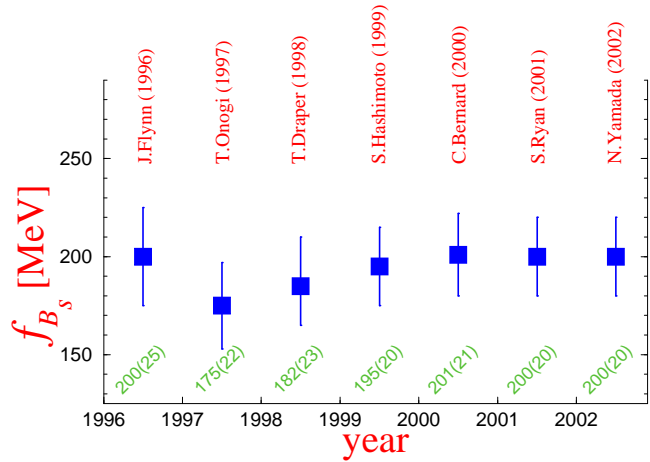}}
\caption[]{World average of the quenched lattice estimates of $F_{B_s}$.}
\end{figure}

In this letter, we present results of
a relativistic calculation of decay constants
in the framework of full Salpeter equation.
The full Salpeter equation is a relativistic equation
describing a bound state. Since this method has a very
solid basis in quantum field theory, it is very good
in describing a bound state which is a relativistic
system. In a previous paper \cite{cskimwang}, we solved the
instantaneous Bethe-Salpeter equation \cite{BS}, which is also
called full Salpeter equation \cite{salp}. After we solved the full
Salpeter equation, we obtained the relativistic wave function of
the bound state. We used this wave function to calculate the
average kinetic energy of the heavy quark inside a heavy meson in
$0^{-}$ state, and obtained values which agree very well
with recent experiments. We also found there that the relativistic
corrections are quite large and cannot be
ignored \cite{cskimwang}. In this letter we use this method
to predict the values of decay constants of heavy mesons in
$0^{-}$ state.

\section{Decay constants of $0^{-}$ State}

In this Section, we will calculate the decay constants of heavy
mesons in $0^{-}$ state by using the full Salpeter method. In the
previous paper\cite{cskimwang}, we wrote the relativistic
wave function of $0^{-}$ state as:
\begin{equation}
\varphi_{^{1}S_0}({\vec q})=\left[
{\not\!P}\varphi_1({\vec q})
+M_{H}\varphi_2({\vec q})-{\not\!q}_{\perp}\varphi_2({\vec q})
\frac{M_{H}(\omega_{_Q}-\omega_{q})}{(m_{q}\omega_{_Q}+m_{_Q}\omega_{q})}+
{\not\!q}_{\perp}{\not\!P}
\varphi_1({\vec q})\frac{(\omega_{_Q}+
\omega_{q})}{(m_{q}\omega_{_Q}+m_{_Q}\omega_{q})}\right]\gamma_{_5}\;.
\end{equation}
where $m_{_Q}$, $m_{_q}$ and $M_H$ are the masses of heavy quark,
light quark, and the corresponding heavy meson, respectively;
$p_{_Q}$ and $p_{q}$  are the momenta of the constituent
quarks, and $P$ the total momentum of the heavy meson.
$q$ is the relative momentum of the
meson defined as
$$q\equiv p_{_Q}-{\alpha}_{1}P\equiv {\alpha}_{2}P-p_{q},$$
where $${\alpha}_{1}\equiv\frac{m_{_Q}}{m_{_Q}+m_{q}}~, \;\;
{\alpha}_{2}\equiv\frac{m_{q}}{m_{_Q}+m_{q}}~;$$ the $\omega_Q$
and $\omega_q$ are defined as
$$\omega_Q \equiv \sqrt{m_Q^2 - q_{\perp}^2}, \;\;
\omega_q \equiv \sqrt{m_q^2 - q_{\perp}^2},$$ where the orthogonal
part $q_{\perp}$ of momentum $q$ is defined as
$$q^{\mu}=q^{\mu}_{\parallel}+q^{\mu}_{\perp}\;,$$
$$q^{\mu}_{\parallel}\equiv (P\cdot q/M^{2}_{H})P^{\mu}\;,\;\;\;
q^{\mu}_{\perp}\equiv q^{\mu}-q^{\mu}_{\parallel}\;.$$ In the
center-of-mass system of the heavy meson, $q_{\parallel}$ and
$q_{\perp}$ turn out to be the usual components
$(q_0, {\vec 0})$ and $(0,{\vec q})$,
and
$\omega_{Q}=(m_{_Q}^{2}+{\vec q}^{2})^{1/2}$ and
$\omega_{q}=(m_{q}^{2}+{\vec q}^{2})^{1/2}$.
Wave functions $\varphi_1({\vec q})$ and
$\varphi_2({\vec q})$ represent the
eigenfunction of the heavy meson obtained by solving the full Salpeter
equation. They will fulfill the normalization condition
\begin{equation}\int\frac{d{\vec q}}{(2\pi)^3}4\varphi_1
({\vec q})\varphi_2({\vec q})M_{H}^2\left\{
\frac{\omega_{_Q}-\omega_{q}}{m_{_Q}-m_{q}}+\frac{m_{_Q}-m_{q}}{\omega_{_Q}-\omega_{q}}
+\frac{2{\vec q}^2(\omega_{_Q}m_{_Q}+\omega_{q}m_{q})}{(\omega_{_Q}m_{q}+\omega_{q}m_{_Q})^2}
\right\}=2M_{H}.
\end{equation}

Several input parameters are needed when we solve the full Salpeter
equation. They are similar to those
in the usual potential model. In Ref.~\cite{cskimwang}, we
obtained the following best fit values of the input parameters
by fitting the mass spectra for heavy mesons of $0^-$ states:
\begin{center}
$a=e=2.7183$, $\alpha=0.06$ GeV, $V_0=-0.60$ GeV,
$\lambda=0.2$ GeV$^2$, $\Lambda_{QCD}=0.26$ GeV~~~ and \\
$m_b=5.224$ GeV, $m_c=1.7553$ GeV, $m_s=0.487$ GeV, $m_d=0.311$
GeV, $m_u=0.305$ GeV.
\end{center}
With this parameter set, we solved the full Salpeter
equation and obtained the eigenvalues and the eigenfunction of the
ground heavy $0^{-}$ states. We will not show here how
the full Salpeter equation is solved and what the
calculated mass spectra are, interested reader can find them in
Ref.~\cite{cskimwang}. We can use the obtained eigenfunction
of heavy mesons to calculate the decay constant $F_P$. The
decay constant is defined as
\begin{eqnarray}
\langle0|\bar{q_1}\gamma_\mu \gamma_5 q_{2} |P\rangle &\equiv& i
F_{P} P_\mu,
\end{eqnarray}
which can be written in the language of the Salpeter wave functions
as:
\begin{eqnarray}
\langle0|\bar{q}\gamma_\mu \gamma_5 Q |P\rangle &=&
i\sqrt{N_c}\int Tr \left[\gamma_\mu
(1-\gamma_5)\varphi_{^{1}S_0}({\vec q})
\frac{d{\vec q}}{(2\pi)^3} \right]=
i4\sqrt{N_c}P_\mu \int
\frac{d {\vec q}}{(2\pi)^3}
\varphi_{1}({\vec q}) \label{fp}.
\end{eqnarray}
Therefore, we have
\begin{eqnarray}
F_{P} = 4\sqrt{N_c} \int
\frac{d{\vec q}}{(2\pi)^3}
\varphi_{1}({\vec q}),\end{eqnarray} and the
calculated values of decay constants are displayed in Table 2.

In Table 3, we show the theoretical uncertainties of our results
for the decay constants. These uncertainties are obtained by
varying all the input parameters simultaneously within 10\% of the
central values, and taking the largest variation of the decay constant.

\begin{table*}[hbt]
\setlength{\tabcolsep}{0.5cm} \caption{\small Decay constants of
heavy $0^{-}$ meson (in GeV) as predicted
by the relativistic Salpeter method}
\label{tab2}
\begin{tabular*}{\textwidth}{@{}c@{\extracolsep{\fill}}cccccccc}
 \hline \hline&$F_{B_c}$&$F_{B_s}$&$F_{B_d}$&$F_{B_u}$
&$F_{\eta_c}$&$F_{D_s}$&$F_{D_d}$&$F_{D_u}$\\ \hline
{\phantom{\Large{l}}}\raisebox{+.2cm}{\phantom{\Large{j}}}
&322&216&197&196&292&248&230&230
\\ \hline\hline
\end{tabular*}
\end{table*}

\begin{table*}[hbt]
\setlength{\tabcolsep}{0.5cm} \caption{\small The theoretical relative
uncertainties, obtained as explained in the text, in per cents ($\%$).}
\begin{tabular*}{\textwidth}{@{}c@{\extracolsep{\fill}}ccccccccc}
 \hline \hline&&${B_c}$&${B_s}$&${B_d}$&${B_u}$
&${\eta_c}$&${D_s}$&${D_d}$&${D_u}$\\ \hline
{\phantom{\Large{l}}}\raisebox{+.2cm}{\phantom{\Large{j}}}
&${\Delta
F_P}/F_P$&$\pm$13&$\pm$15&$\pm$15&$\pm$15&$\pm$8.6&$\pm$11&$\pm$11&$\pm$11
\\ \hline\hline
\end{tabular*}
\end{table*}

\begin{table*}[hbt]
\setlength{\tabcolsep}{0.5cm} \caption{\small Recent calculations
by other methods.
Here PM means Potential Model, BS means
Bethe-Salpeter method, QL means Quenched Lattice, AQL means
Average Quenched Lattice, UL means Unquenched Lattice, AUL means
Averaged Unquenched Lattice, QSR means QCD Sum Rules.
In Ref.~\cite{bernard}, the uncertainties are statistical, systematic
within the $N_f=2$ partially quenched approximation, the
systematic errors due to partial quenching and the missing virtual
strange quark, and an estimate of the effect of chiral logarithms,
respectively. In Ref. \cite{yamada}, the uncertainties are from
statistics, chiral extrapolation and systematics.}
\begin{tabular*}{\textwidth}{@{}c@{\extracolsep{\fill}}ccccc}
\hline \hline Ref.&$F_{B_s}$&$F_{B_d}$ or $F_{B_u}$
&$F_{D_s}$&$F_{D_d}$ or $F_{D_u}$\\ \hline
{\phantom{\Large{l}}}\raisebox{+.2cm}{\phantom{\Large{j}}}
PM\cite{ebert} &196$\pm$20&178$\pm$15&266$\pm$25&243$\pm$25
\\
{\phantom{\Large{l}}}\raisebox{+.2cm}{\phantom{\Large{j}}}
BS\cite{wangzg}&&192&&
\\ \\{\phantom{\Large{l}}}\raisebox{+.2cm}{\phantom{\Large{j}}}
QL\cite{bernard} &{\small
217(6)($^{+32}_{-28}$)($^{+9}_{-3}$)($^{+17}_{-0}$)}& {\small
190(7)($^{+24}_{-17}$)($^{+11}_{-2}$)($^{+8}_{-0}$)}&{\small
241(5)($^{+27}_{-26}$)($^{+9}_{-4}$)($^{+5}_{-0}$)} &{\small
215(6)($^{+16}_{-15}$)($^{+8}_{-3}$)($^{+4}_{-0}$)}
  \\
  {\phantom{\Large{l}}}\raisebox{+.2cm}{\phantom{\Large{j}}}
 QL\cite{juttner} &&&252$\pm$9&\\
{\phantom{\Large{l}}}\raisebox{+.2cm}{\phantom{\Large{j}}}
 AQL\cite{ryan}
&200$\pm$20&173$\pm$23&230$\pm$14&203$\pm$14\\
{\phantom{\Large{l}}}\raisebox{+.2cm}{\phantom{\Large{j}}}
 UL\cite{yamada}&&190(14)(07)(19)&&\\

{\phantom{\Large{l}}}\raisebox{+.2cm}{\phantom{\Large{j}}}
 AUL\cite{ryan}&230$\pm$30&198$\pm$30&250$\pm$30&226$\pm$15\\ \\

{\phantom{\Large{l}}}\raisebox{+.2cm}{\phantom{\Large{j}}}
 QSR\cite{narison}&236$\pm$30&203$\pm$23&235$\pm$24&204$\pm$20\\
{\phantom{\Large{l}}}\raisebox{+.2cm}{\phantom{\Large{j}}}
 QSR\cite{penin}&&206$\pm$20&&195$\pm$20
\\
\hline\hline
\end{tabular*}
\end{table*}

In Table 4, for comparison, we show recent theoretical predictions
for the decay constants as obtained by other methods. For example,
we display the recent values from relativistic potential model
(PM) \cite{ebert} based on the quasi-potential approach; most
recent value of $F_{B}$ from another version of using
Bethe-Salpeter method (BS) \cite{wangzg}, which is also a
relativistic result; recent values from the averaged lattice
results both in quenched (AQL) and unquenched (AUL) approximation
\cite{ryan}; most recent values from quenched lattice (QL) QCD
\cite{bernard,juttner} and unquenched lattice (UL) QCD
\cite{yamada}; and values from QCD Sum Rules (QSR) \cite{narison,
penin}. As can be seen from Tables 2 and 4, our values of the
decay constants by solving the Salpeter equation, agree with these
recent results by other methods. In particular, they agree very
well with the recent average of the unquenched lattice QCD (AUL)
\cite{ryan}.
Our value $F_{D_s} \approx~ 248$ GeV is smaller than the most
recent experimental central value, the ALEPH's value $F_{D_s}
\approx~ 285\pm 19\pm 40$, but still within the
experimental uncertainties.

\begin{table*}[hbt]
\setlength{\tabcolsep}{0.5cm} \caption{\small Ratios
$F_{B_s}/F_{B_d}$, $F_{D_s}/F_{D_d}$ and the Grinstein
ratio by this work and by other methods. In Ref.~\cite{onogi}, the
first and second uncertainty are the statistical and the systematic
errors. }
\begin{tabular*}{\textwidth}{@{}c@{\extracolsep{\fill}}cccc}
 \hline \hline &Ref.&$\frac{F_{B_s}}{F_{B_d}}$&$\frac{F_{D_s}}{F_{D_d}}$
&R1\\ \hline
{\phantom{\Large{l}}}\raisebox{+.2cm}{\phantom{\Large{j}}}
&This work&1.10$\pm$0.01&1.08$\pm$0.01&1.02$\pm$0.02\\
 {\phantom{\Large{l}}}\raisebox{+.2cm}{\phantom{\Large{j}}}
&PM\cite{ebert}&1.10$\pm$0.21&1.09$\pm$0.22&1.01$\pm$0.40\\ \\
{\phantom{\Large{l}}}\raisebox{+.2cm}{\phantom{\Large{j}}}
&QL\cite{bernard}&1.16(1)(2)(2)($^{+4}_{-0}$)&
1.14(1)($^{+2}_{-3}$)(3)(1)&1.02(2)(4)(4)($^{+4}_{-1}$)\\
{\phantom{\Large{l}}}\raisebox{+.2cm}{\phantom{\Large{j}}}
&UL\cite{onogi}&&&1.018$\pm$0.006$\pm$0.010\\
 {\phantom{\Large{l}}}\raisebox{+.2cm}{\phantom{\Large{j}}}
&AQL\cite{ryan}&1.15$\pm$0.03&1.12$\pm$0.02&1.03$\pm$0.05\\
{\phantom{\Large{l}}}\raisebox{+.2cm}{\phantom{\Large{j}}}
&AUL\cite{ryan}&1.16$\pm$0.05&1.12$\pm$0.04&1.04$\pm$0.08\\
\\ {\phantom{\Large{l}}}\raisebox{+.2cm}{\phantom{\Large{j}}}
&QSR\cite{narison2}&1.16$\pm$0.05&1.15$\pm$0.04&1.01$\pm$0.08
\\ \hline\hline
\end{tabular*}
\end{table*}

There are other interesting quantities such as the ratios of decay
constant $F_{B_s}/F_{B_d}$, $F_{D_s}/F_{D_d}$, and
the Grinstein ratio \cite{grinstein} defined as
\begin{eqnarray}
R1=(\frac{F_{B_s}}{F_{B_d}})/(\frac{F_{D_s}}{F_{D_d}}),\end{eqnarray}
which is a quantity sensitive to the light quark mass $m_s$. In
Table 5 we show our values of these ratios and some values
obtained by other methods in recent literature. Our uncertainties
come from the aforementioned 10 per cent changes of the
parameters. The uncertainties of the ratios of decay constants of
Ref.~\cite{ebert} are large. This is so because the authors of
Ref.~\cite{ebert} did not give the uncertainties for these ratios.
We estimated the uncertainties of these ratios on the basis of
their given uncertainties of the decay constants. In the same way
we estimated the uncertainties of the Grinstein ratio of other
references shown in Table 5, with the exception of those of
Ref.~\cite{onogi}. {}From Table 5 one can see that our values of
ratios $F_{B_s}/F_{B_d}$ and $F_{D_s}/F_{D_d}$ agree with these
recent theoretical results. In particular, our central values are
very close to those of the relativistic potential model
\cite{ebert},
%smaller
%than the results of lattice QCD\cite{bernard,ryan} and QCD Sum
%Rules\cite{narison}.
and our central value of the Grinstein ratio $R1=1.02$
agrees well with the values estimated by other methods.

In conclusion, we calculated the decay constants of heavy $0^{-}$
mesons by means of the relativistic Salpeter method. We obtained
$F_{D_s} \approx~ 248~\pm ~27~$, $F_{D} \approx~ 230~\pm ~25~
(D^0,D^\pm)$, $F_{B_s} \approx~ 216~\pm ~32~$, $F_{B} \approx~
196~\pm ~29~ (B^0,B^\pm)$, $F_{B_c} \approx~ 322~\pm ~42~$ and
$F_{\eta_c} \approx~ 292~ \pm ~25~$ MeV.\\

The work of C.S.K. was supported in part by CHEP-SRC
Program, in part by Grant No. R02-2003-000-10050-0 from BRP of the
KOSEF. The work of G.W. was supported by BK21 Program. The work
of G.C. was supported by FONDECYT (Chile) grant no.~1010094.
\\

\end{document}